\documentclass[prd,aps,amsfonts,eqsecnum,superscriptaddress,nofootinbib,longbibliography,notitlepage]{revtex4-1}

\usepackage{preamble}

\begin{document}

\title{Effective field theory for ersatz Fermi liquids}

\author{Xiaoyang Huang}
 \email{xiaoyang.huang@colorado.edu}
\affiliation{Department of Physics and Center for Theory of Quantum Matter, University of Colorado, Boulder CO 80309, USA}
\affiliation{Kavli Institute for Theoretical Physics, University of California, Santa Barbara, CA 93106, USA}

\author{Andrew Lucas}
 \email{andrew.j.lucas@colorado.edu}
\affiliation{Department of Physics and Center for Theory of Quantum Matter, University of Colorado, Boulder CO 80309, USA}

\author{Umang Mehta}
 \email{umang.mehta@colorado.edu}
\affiliation{Department of Physics and Center for Theory of Quantum Matter, University of Colorado, Boulder CO 80309, USA}

\author{Marvin Qi}
 \email{marvin.qi@colorado.edu}
\affiliation{Department of Physics and Center for Theory of Quantum Matter, University of Colorado, Boulder CO 80309, USA}

\begin{abstract}
     We apply ``hydrodynamic" effective field theory techniques to an ersatz Fermi liquid.  Our effective theory, which captures the correlation functions of density operators at each angle on the Fermi surface, can only deviate from conventional Fermi liquid behavior if the effective theory is non-local. Neglecting non-local effects, the ersatz Fermi liquid's effective action is the Legendre transform of the effective action for Fermi liquids, based on the coadjoint orbit method, up to irrelevant corrections. 
\end{abstract}

\date{\today}

\maketitle 

\section{Introduction}
Ordinary metals are well-described by Landau's Fermi liquid (FL) theory \cite{Landau:1956zuh, Benfatto:1990zz, Polchinski:1992ed, Shankar:1993pf}.  The experimental observation of physics not apparently compatible with this description, such as $T$-linear resistivity \cite{keimer2015highTc, phillips2022str} (at very low temperatures), incoherent plasmons \cite{Mitrano_2018,PhysRevX.9.041062}, and rapid quasiparticle decay rates \cite{heartnoll2022planck}, suggests the need for a new theory of a non-Fermi liquid (NFL): a strongly correlated metal without quasiparticles.  Various approaches have attempted to tackle this problem, primarily focused on quantum criticality or singular interactions in metals (see, e.g. \cite{Lee:2017njh} and references therein) and disorder-driven strangeness \cite{sachdev1993syk, sachdev2022statistical, esterlis2021largeN, guo2021largeN, patel2023largeN}.

A series of recent papers \cite{Else_PRX_EFL, Else2021strange, Else_Senthil_criticaldrag, Senthil_gift, shi2023loop,Else_2023,Else:2023xgk} has argued, in the spirit of an effective field theory (EFT), that a large class of metals have an infinite-dimensional emergent symmetry group at their IR fixed point.   The resulting phase was dubbed the ``ersatz Fermi liquid" (EFL).  For a single component Fermi surface in two spatial dimensions, the EFL's emergent symmetry group is LU(1) (which we will review shortly).  If such an emergent symmetry group is indeed present, it should be an ingredient into future effective field theories of non-Fermi liquid metals, and has at least appeared missing in many previous constructions.  For example, Else has recently argued \cite{Else_2023} that LU(1) symmetry is powerful enough to lead to a ``conventional"-looking theory of zero sound (at least, before considering dissipation), which is, at least in principle, a clear prediction of the EFL EFT.  (In practice, since disorder or irrelevant interactions could break LU(1), such predictions may be hard to find in experiment.  Indeed, the present paper is mostly an effort to better understand the theoretical implications of the EFL theory.)

On the other hand, one of us has recently demonstrated that LU(1) is  the linear approximation to a non-Abelian group that generates the low energy space of states and hence controls the EFT of a \emph{conventional} Fermi liquid \cite{Delacretaz:2022coadj, Mehta2023postmodern} (see \cite{haldane2005luttingers, CastroNetoFradkin:1994, Houghton:2000bn,Khveshchenko_1994} for previous approaches to bosonizing Fermi surfaces in $d\ge 2$).   Since the nonlinear corrections to this bosonized theory for even an ordinary Fermi liquid break LU(1), it is thus unclear if LU(1) should be imposed as a fundamental emergent symmetry of a more general NFL, especially when trying to build a controlled EFT description of such a NFL.

In this paper, we take the perspective that it is worth unpacking, from an effective field theorist's point of view, what some of the implications must be of the EFL proposal.  We thus present a discussion of the EFT of an EFL using the formal methods developed for fluids and superfluids over the past two decades \cite{Crossley:2015evo, Glorioso:2017fpd, Glorioso:2017lcn, Dubovsky:2011sj, Jensen:2012jh,Delacretaz_2020} (see \cite{glorioso2018lectures} for a review).  Such methods are, in fact, quite natural for this problem: after all, a theory of a superfluid/fluid is simply the EFT of a theory with U(1) symmetry which is/is not spontaneously broken.  Our goal is to sharply pinpoint where, if at all, the most general possible EFT of an EFL can differ from the conventional FL theory.  While such tasks are inevitably complicated by the quite subtle nature of renormalization group analyses for theories with a Fermi surface \cite{Benfatto:1990zz, Shankar:1993pf, Polchinski:1992ed, Altshuler1994patchNFL, Nayak:1993uh, Nayak:1994ng, Metlitski:2010pd, MetlitskiL2010sdw, Lee:2009epi, mandal2015uv/ir_nfl, ye2022uv/ir_mfl}, we will argue that so long as the only slow/gapless degrees of freedom in the theory are the bosons associated with the LU(1) symmetry (physically, related to the fermion number at each angle on the Fermi surface),  our EFT either appears inconsistent, or consistent with usual FL theory with \emph{very irrelevant} interactions.  In other words, EFLs on their own are so strongly constrained that it is not possible for them to behave differently from ordinary FLs in the IR.  The exception to this conclusion is when the EFL is coupled to another gapless mode, such as a critical boson, in an LU(1)-invariant manner, in which case the EFL EFT may become strongly correlated; we do not attempt a careful analysis of such a regime in this work, besides to note that such an EFT will necessarily appear non-local. 


Using the hydrodynamic interpretation of the EFL EFT, we are also able to make a physically transparent connection to the coadjoint orbit formalism of \cite{Delacretaz:2022coadj}.   Up to irrelevant corrections, the EFL EFT we obtain is simply the Legendre transform of the coadjoint orbit theory (written in an alternative way). This simple relationship is, in fact, required for consistency with thermodynamic requirements on the hydrodynamic effective field theories for fluids.

In this paper, we focus only on EFTs in flat spacetime in $d=2$; however, we note that our approach is the natural starting point for coupling an EFL to general backgrounds, including curved space.   We also neglect the higher-dimensional setting, although our conclusions do not sensitively depend on the dimensionality of space.

\section{Symmetries of the ersatz Fermi liquid}\label{sec:symmetry}

We set out to construct an effective field theory for a fluid charged under an LU(1) symmetry. An element of $\mathrm{LU}(1)$ can be presented as a function $\alpha: \mathrm{U}(1) \to  \mathrm{U}(1)$; here we are thinking of $\mathrm{U}(1) = \mathbb{R}/2\pi\mathbb{Z}$; the group operation is simply addition of functions, i.e. LU(1) is Abelian.
In an effective field theory for a system with an (unbroken) LU(1) symmetry, following \cite{Dubovsky:2011sj, Crossley:2015evo, Glorioso:2017fpd}, we anticipate the slow degrees of freedom to include phase variables associated with each generator of LU(1).  Letting $\theta \in \mathbb{R}/2\pi\mathbb{Z}$ denote the parameter of LU(1) --in more physical terms, the angle along the Fermi surface -- this implies we have degrees of freedom \begin{equation} \label{eq:lu1 shift}
    \varphi(x^\mu, \theta) \rightarrow \varphi(x^\mu, \theta) + \alpha(\theta)
\end{equation}
that shift nonlinearly under LU(1) transformations.   Here $x^\mu=(t,x^i)$ includes both time and space coordinates: as is usual $\mu\nu$ denote spacetime indices, while $ij$ denote space indices alone.

We assume that there are no further degrees of freedom in the theory.  From a hydrodynamic perspective this should be reasonable: the only other slow degrees of freedom not accounted for could be the energy and momentum.   However, to the extent that the energy and momentum currents have vacuum expectation values (as this paper focuses on zero temperature physics), these overlaps are expected to come entirely from the presence of the LU(1) conserved charges.  For this reason, we will build an effective Lagrangian $\mathcal{L}(\varphi)$ alone.  The LU(1) conserved charge densities will be denoted as $\rho(x^\mu, \theta)$, and can be obtained from Noether's Theorem straightforwardly: \begin{equation}\label{eq:noether}
    \rho(x^\mu,\theta) = \frac{\partial \mathcal{L}}{\partial (\partial_t \varphi(x^\mu,\theta))}.
\end{equation} 

In order to ensure that the LU(1) symmetry is not spontaneously broken, we demand that the theory be invariant under the following ``reparameterization" symmetry: \begin{align}\label{eq:reparametrization}
    \varphi \rightarrow \varphi + \chi(x^i, \theta),
\end{align}
where $\chi$ is a time-independent function from spatial coordinates to LU(1).
We note that the need for such a reparameterization symmetry has been a somewhat ``confusing" aspect of hydrodynamic EFTs -- at least insofar as it's physical origin is not clear.  When we compare to the coadjoint orbit formalism, we will see at least one interpretation for this symmetry.

We can also couple the EFT to background LU(1) gauge fields. As we have emphasized above, LU(1) is an Abelian group; its gauge fields therefore look like conventional U(1) gauge fields, labeled by an additional ``flavor'' $\theta$: $A_\mu(x^\mu,\theta)$, which transform under the gauge transformations as
\begin{align}\label{eq:gauge transformation}
    A_\mu(x^\mu,\theta) \rightarrow A_\mu(x^\mu,\theta) - \p_\mu \lambda(x^\mu,\theta).
\end{align}
Since the same gauge transformation would transform
\begin{align}
    \varphi(x^\mu,\theta) \rightarrow \varphi(x^\mu,\theta) + \lambda(x^\mu,\theta),
\end{align}
gauge-invariant objects are given by
\begin{align}
    B_\mu \equiv \p_\mu \varphi + A_\mu.
\end{align}
Not all of them will be our invariant building blocks. In fact, because the reparametrization symmetry \eqref{eq:reparametrization} is \emph{not} a special case of the gauge-transformation above ($\chi$ does not modify $A_\mu$), we forbid the spatial components to appear at the leading order in EFT.  The timelike component
\begin{align}
    B_{t}(x^\mu,\theta) \equiv \mu(x^\mu,\theta)
\end{align}
is the chemical potential for the LU(1) conserved charges; notice that there is one for every $\theta$. Other invariant blocks can be obtained by acting spatial and temporal derivatives and they should be organized by the derivative expansion; $\theta$-derivative, however, is allowed to be added on invariant blocks without raising its derivative orders.\footnote{One should be cautious that as soon as $\theta$ acquires a nonzero scaling dimension, the counting changes.  We will return to this point later.}

Besides LU(1), we assume an SO(2) rotational symmetry that not only acts on the spatial manifold, but also transforms $\theta$. Unlike the usual fluid, LU(1) contains a vector-like object under SO(2) given by\footnote{We note that such a vector is interpreted in the original EFL paper of  \cite{Else_PRX_EFL,Else:2023xgk} as the field strength $\epsilon^{ij}F_{i\theta} \equiv \epsilon^{ij}(\p_i A_\theta - \p_\theta A_i) = - \epsilon^{ij}\p_\theta n_i = n^j$ where they set $\p_i A_\theta = 0$ and $A_i = n_i$. \label{footnote field strength}}
\begin{align}\label{eq:n}
    \vn(\theta) = (\cos\theta, \sin\theta)^T,
\end{align}
which denotes the direction of the Fermi momentum/velocity (for a circular Fermi surface) as a function of $\theta$.  From the perspective of the EFT, this vector can be incorporated into invariant building blocks and into the Lagrangian.  


A particularly critical place where this will appear is in the description of an LU(1) anomaly, which encapsulates the earlier result that \cite{Fradkin_prb,Fradkin_prl,haldane2005luttingers}
\begin{align}\label{eq:KM algebra}
    [\rho(t,x_i,\theta),\rho(t,x_i',\theta')] = \ii C \vn \cdot \nabla \delta^{(2)}(x_i-x'_i)\delta(\theta - \theta')
\end{align}
with a constant $C$.
This commutation relation gives a central extension of the Abelian LU(1) algebra. The physical consequence of this generalized algebra is that under a static perturbation \begin{equation}\label{eq:Delta H}
    \Delta H = -\int \ud t \ud^2 x \ud\theta~ V(x_i) \rho(x^\mu,\theta),
\end{equation} 
the charge density is no longer conserved and is broken by the electric field $\vect E = - \nabla V$:
\begin{align}\label{eq:anomalous conservation}
    \p_t \rho(x^\mu,\theta) = \ii [\Delta H,\rho(x^\mu,\theta)] = C\vn\cdot\vect E.
\end{align} 
In analogy to the Luttinger liquid, the anomalous charge conservation renders the filling $\nu$ to be fixed by the volume of the Fermi sea $\mathcal{V}_{\F}$: $\nu = \abs{C} \mathcal{V}_{\F}/p_\F$, which is the Luttinger's theorem when $\abs{C} = p_\F/(2\pi)^2$\cite{Else_PRX_EFL,Wen_prb_2021}; we will see later when comparing to the coadjoint orbit theory that this is indeed the case.

Intriguingly, if the Fermi surface has lower spatial symmetry, the anomaly can be generalized. For example, for a rectangular Fermi surface with reflection and two-fold rotation, we can have $\p_t \rho = h_{ij}n^i E^j$ where $h_{ij}$ is an invariant tensor. We shall leave a study of the generic anomaly for future work, and focus on the isotropic Fermi surface in this paper.

In \cite{Else_PRX_EFL}, an extra gauge field component $A_\theta$ is introduced.  Introducing this extra gauge field component enables a simple interpretation of the anomaly discussed above in terms of a single anomaly coefficient. At the same time, these authors demanded $A_\theta$ to vanish at linear order in order to not generate responses in the absence of the magnetic field, which essentially forbids $A_\theta$ to couple to the dynamical degrees of freedom. This condition raises the question of whether to include $A_\theta$ in an EFT Lagrangian.  We argue that it is not natural to incorporate $A_\theta$, in part because it is not clear whether the corresponding operator it would source has a physical interpretation. (In contrast, $A_\mu(\theta)$ sources $J^\mu(\theta)$, the conserved current associated with each LU(1) conservation law.)  Moreover, since there is only spacetime diffeomorphism, $A_\theta$ behaves like a scalar rather than a 1-form, and thus is distinct from the U(1) gauge field $A_\mu$. Therefore, we do not couple our EFT to $A_\theta$, and for this reason we did not discuss $A_\theta$ when describing the LU(1) anomaly.

We will not include momentum conservation in our hydrodynamic EFT. However, unlike in usual kinetic theory we do not then also explicitly break the LU(1).   In ordinary kinetic theory this arises through impurity scattering in the collision integral, but our EFT is dissipationless and at zero temperature.  

\textcolor{red}{Lastly, given the application of the LU(1)-invariant EFT to metallic physics, it might be natural to consider the possibility of an instability to superconductivity.  This would require introducing additional bosonic degrees of freedom for such ``cooperons" \cite{Delacretaz:2022coadj,Mehta2023postmodern}, and we do not do so in this paper; our goal is to understand the EFL.}

\section{Lagrangian effective field theory}\label{sec:eft}
Having discussed LU(1) symmetry and the degrees of freedom for our EFT, it remains to actually build the Lagrangian.  To do so at the ideal fluid level, it is useful to think of the fluid Lagrangian as defined in such a way that it reproduces the generating function of correlation functions in the ground state of the system:
\begin{align}
    \mathrm{e}^{W[A_\mu]} = \left\langle \mathrm{e}^{\ii \int\ud t\ud^2x \mathrm{d}\theta  ~ A_\mu(\theta) J^\mu(\theta) } \right\rangle  = \int \mathrm{D}\varphi ~\mathrm{e}^{\ii \int\ud t\ud^2x  ~ \mL [\varphi;A_\mu] }.
\end{align}
We now find the effective Lagrangian $\mL$ that satisfies the symmetries discussed in \secref{sec:symmetry}. At the lowest order, we have 
\begin{align}\label{eq:L0}
    \mL_0 = P(\mu,\ldots),
\end{align}
where $P$ is the thermodynamic pressure \cite{Dubovsky:2011sj,Jensen:2012jh}, and $\ldots$ includes either $\partial_t B_i$, $\partial_i B_t$, $\partial_\theta B_t$, etc. Taking the system to be in thermal equilibrium at chemical potential $\mu_0(\theta)$, we can expand the Lagrangian in perturbations $\delta\mu\equiv \mu - \mu_0$ as
\begin{align}\label{eq:L0 EFT}
    \mL_0 \approx \frac{1}{2}\int \ud\theta \ud \theta' ~ g_2(\theta,\theta') \delta \mu(x^\mu,\theta) \delta \mu(x^\mu,\theta')+\frac{1}{3!}\int \ud\theta \ud \theta'\ud \theta'' ~ g_3(\theta,\theta',\theta'') \delta \mu(x^\mu,\theta) \delta \mu(x^\mu,\theta')\delta \mu(x^\mu,\theta'')+\ldots,
\end{align}
where the dots include higher order in $\delta\mu$ organized by the function $g_{n>3}$. Notice that the functionals $g_n(\theta,\ldots)$ are in principle completely non-local in $\theta$ and can contain $\theta$-derivatives acting on $\delta \mu$. 
According to the first law of thermodynamics, the charge density is given by
\begin{align}\label{eq:charge density}
    \rho(x^\mu,\theta) \equiv \frac{\p P}{\p\mu}\approx \int \ud \theta' ~ g_2(\theta,\theta')  \delta \mu(x^\mu,\theta') + \ldots.
\end{align}

Due to the anomalous charge conservation \eqref{eq:anomalous conservation}, the generating function in the presence of the background gauge fields is not invariant under a small gauge transformation; it thus has  a \emph{local LU(1) anomaly}. In particular, \eqref{eq:anomalous conservation} implies that\footnote{ The anomalous equation \eqref{eq:anomalous conservation} is written in terms of covariant currents $J^\mu$ which are gauge-invariant. The gauge non-invariance of the generating function is related to the anomalous equation $\p_\mu \tilde J^\mu = \frac{1}{2} C \vect n\cdot \vect E$ associated with the consistent current $\tilde J^{t} = J^t + \frac{1}{2} C \vect n \cdot A $ and $\tilde J^{i} = J^i- \frac{1}{2} C n^i A_t $.}
\begin{align}\label{eq:W anomaly}
    -\ii W[A_\mu-\p_\mu \lambda] = -\ii W[A_\mu] +\frac{C}{2} \int\ud t\ud^2 x \ud \theta ~\lambda n^i F_{it},
\end{align}
where $F_{\mu\nu} = \p_\mu A_\nu - \p_\nu A_\mu$ and $E_i = F_{it}$. The minimal Lagrangian that satisfies \eqref{eq:W anomaly} is given by
\begin{align}\label{eq:Lanom}
    \mL_{\mathrm{anom}} =\frac{C}{2} \int \ud\theta ~ \varphi n^i F_{it}.
\end{align}
However, $\mL_{\mathrm{anom}}$ is not invariant under the reparametrization symmetry \eqref{eq:reparametrization}. To make the total action invariant, we write 
\begin{align}
    \mL = \mL_0 + \mL_{\mathrm{inv}}+\mL_{\mathrm{anom}},
\end{align}
where $\mL_0$ and $\mL_{\mathrm{anom}}$ are given by \eqref{eq:L0} and \eqref{eq:Lanom}, and $\mL_{\mathrm{inv}}$ is the additional part that is gauge-invariant. Choosing 
\begin{align}\label{eq:Linv}
    \mL_{\mathrm{inv}} = \frac{C}{2}\int\ud\theta ~ n^i B_{i} B_t = \frac{C}{2}\int\ud\theta ~ n^i (\p_i \varphi + A_i) (\p_t \varphi + A_t) ,
\end{align}
we find that (after subtracting off total derivatives) the total $\mL$ is invariant under the reparametrization symmetry. Collecting \eqref{eq:L0}, \eqref{eq:Lanom} and \eqref{eq:Linv}, we arrive at
\begin{align}\label{eq:Ltot}
    \mL = P(\mu) + \frac{C}{2} \int \ud\theta ~ n^i (\p_i \varphi + A_i) \mu + \frac{C}{2} \int \ud\theta ~ \varphi n^i F_{it}.
\end{align}
This Lagrangian (and the symmetries that underlie the effective theory) appear different from those put forth in \cite{Wen_prb_2021}; the construction of \cite{Wen_prb_2021} does not impose the reparameterization symmetry (\ref{eq:reparametrization}), which suggests that LU(1) is spontaneously broken (from the hydrodynamic EFT perspective).

By varying with respect to $\varphi$, we obtain the equation of motion
\begin{align}\label{eq:eom main}
    \p_t \rho(x^\mu,\theta) + C \vn \cdot\nabla \delta \mu(x^\mu,\theta) = C \vn \cdot \vect E.
\end{align}
This equation of motion is exact at the ideal hydordynamic level.  Nonlinear terms are, of course, allowed in the expression of $\delta\mu$ as a function of $\rho$. 
In the limit where the EFL describes a Landau's FL, we may write
\begin{align}
    \delta \mu(x^\mu,\theta) \approx \frac{v_\F}{C} \rho(x^\mu,\theta) + \frac{v_\F}{C}\int \ud\theta' ~ F(\theta,\theta')\rho(x^\mu,\theta') + \cdots,
\end{align}
where the first term is due to the approximate linear dispersion near Fermi surface and we have parametrized interacting effects through the Landau parameter $F(\theta,\theta') =F(\theta',\theta) $.  With rotational symmetry one finds $F(\theta,\theta') =F(\theta-\theta') $. Comparing to \eqref{eq:charge density}, we find
\begin{align}\label{eq:g2}
    g_2^{-1}(\theta,\theta') = \frac{v_\F}{C}\left[\delta(\theta - \theta') + F(\theta,\theta')\right],
\end{align}
and the resulting equation of motion \eqref{eq:eom main} is the Boltzmann equation.

Having constructed the most general LU(1) EFT, we now discuss the scaling dimensions of the operators within the EFT to determine the stability of our fixed point $\mu=\mu_0$. To determine the zero-order scaling dimension of the dynamical field $\varphi$, we look at the Gaussian action
\begin{align}\label{eq:S 2}
    S^{(2)}  = \frac{1}{2} \int\ud t \ud^2 x \ud\theta \ud \theta' ~ g_2(\theta,\theta') \p_t\varphi(x^\mu,\theta) \p_t\varphi(x^\mu,\theta') + \frac{C}{2} \int\ud t \ud^2 x \ud\theta~ \vn\cdot \nabla \varphi(x^\mu,\theta) \p_t \varphi(x^\mu,\theta).
\end{align}
Let us divide the spatial directions into $q_\parallel\equiv \vn \cdot \vq$ and $q_\perp \equiv \epsilon^{ij} n_i q_j$. First, the scaling dimensions $[\omega] = [q_\parallel] = 1$ are fixed by the local LU(1) anomaly (we also take $[q_\parallel]=1$ as a matter of convention): at each angle $\theta$, the action \eqref{eq:S 2} describes a $1+1$d Luttinger Liquid propagating along $\vn(\theta)$.  Second, the transverse direction $\epsilon^{ij}n_j$ does not enter the Gaussian action \eqref{eq:S 2}, therefore, we are allowed to choose $[q_\perp] = \alpha\geq 0$ to be an arbitrary number.\footnote{We do not allow for $\alpha<0$, as if this were the case,  there are an infinite number of increasingly relevant operators, suggesting that the effective field theory is sick.  Hence we do not consider this possibility to be compatible with an LU(1) EFT, at least of the kind described in this paper.} \textcolor{red}{We remind the reader that the fact that $q_\parallel$ and $q_\perp$ scale differently is a general feature of field theory in the presence of a Fermi surface \cite{Polchinski:1992ed,Shankar:1993pf}.}  Further, we assume that $[\theta]=0$: if $[\theta]<0$, the Landau parameter would be more relevant than the non-interacting term in \eqref{eq:g2}, and if $[\theta]>0$, terms with more $\theta$-derivatives would be more relevant, and they both suggest an ill-defined EFT. Then, we have
\begin{align}
    [\varphi] = \frac{\alpha}{2}.
\end{align}
The leading nonlinear correction is the cubic action 
\begin{align}\label{eq:S 3}
    S^{(3)}  = \frac{1}{3!} \int\ud t \ud^2 x \ud\theta \ud \theta' \ud \theta'' ~ g_3(\theta,\theta',\theta'') \p_t\varphi(x^\mu,\theta) \p_t\varphi(x^\mu,\theta')\p_t\varphi(x^\mu,\theta'') .
\end{align}
The scaling dimension of its coefficient $g_3$ is given by
\begin{align}\label{eq:g3 scaling}
    [g_3]  =-1 - \frac{\alpha}{2} <0,
\end{align}
meaning that it is irrelevant. Since all corrections to $\mathcal{L}$ must be reparameterization-invariant and thus depend on $(\partial_t \varphi)^n$ with $n>2$, there is no other spatially local (in $x^\mu$) operator that could be relevant. Therefore, the Gaussian fixed point governed by \eqref{eq:S 2} is stable.

We can infer from the irrelevant nonlinear action the typical decay rate of our LU(1) system according to LU(1)-symmetric decay channels. The fastest decay rate $\Gamma$ is determined by the leading nonlinear correction given in \eqref{eq:S 3}. From Fermi's Golden rule and \eqref{eq:g3 scaling}, we find 
\begin{align}
    \Gamma \sim g_3^2~ \omega^{3+\alpha}.
\end{align}
Note that when estimating $\Gamma$, we have neglected the possibility of UV-IR mixing.  From the EFL EFT perspective, this seems like a reasonable assumption.
We highlight that this decay rate is qualitatively slower than the \emph{ordinary} 
$\omega^2$ decay rate of the local density $\rho(x^\mu,\theta)$, \textcolor{red}{or equivalently the quasi-particle lifetime},
in a Fermi liquid, which arises due to two-body scattering on a Fermi surface.  This two-body scattering is an LU(1)-breaking irrelevant perturbation, \textcolor{red}{and it cannot be written down within our EFT.}\footnote{\textcolor{red}{One heuristic way to understand this point is that $\partial_t \varphi(\theta) \sim \psi^\dagger_\theta\psi_\theta$ corresponds to a density operator for the fermions at angle $\theta$ on the Fermi surface.  The interaction that leads to $\Gamma \sim \omega^2$ in a conventional Fermi liquid includes coefficients such as $\psi^\dagger_{\theta_2+\pi}\psi^\dagger_{\theta_2}\psi_{\theta_1+\pi}\psi_{\theta_1}$, which does not conserve the number of particles at each angle $\theta$.}}   Thus, at least for ordinary Fermi liquids, the LU(1)-invariant EFT does not capture the leading irrelevant corrections to the free (Boltzmann) IR fixed point.  For systems with simply-connected and convex Fermi surfaces in two-spatial dimensions, in the so-called ``tomographic regime" \cite{levitov2019} the dominant decay rates scale faster than $\omega^3$, so it is possible that in these regimes such LU(1)-invariant perturbations could become the dominant corrections.

We then conclude that if the LU(1) charges are the only slow degrees of freedom in the EFL EFT, then the EFL EFT is \emph{identical} to Boltzmann transport.  This is not just a statement about linear response, as was already highlighted in \cite{Else_2023}, but rather a statement about the effective field theory as a whole.  It is not possible, in particular, to find non-Fermi-liquid (larger than $\omega^2$) decay rates for excitations.   

We remark that this is not in contradiction to the $1+1$d Luttinger liquid theory where changing the coefficient of the Gaussian theory will lead to non-Fermi liquid phase. This is because the typical lifetime in a Luttinger liquid is $O(\omega)$ larger than $\omega^2$ \cite{LL_decayrate}.


These statements do not preclude the possibility of coupling the EFL to additional gapless modes, and we will briefly discuss this possibility in Section \ref{sec:nfl}.  However, we do emphasize that without such additional gapless modes, the EFL EFT \emph{must} be identical to conventional kinetic theory up to irrelevant corrections.

\section{Comparison to the coadjoint orbit method}\label{sec:coadjoint orbit}

\begin{figure}[t]
\includegraphics[width=.5 \linewidth]{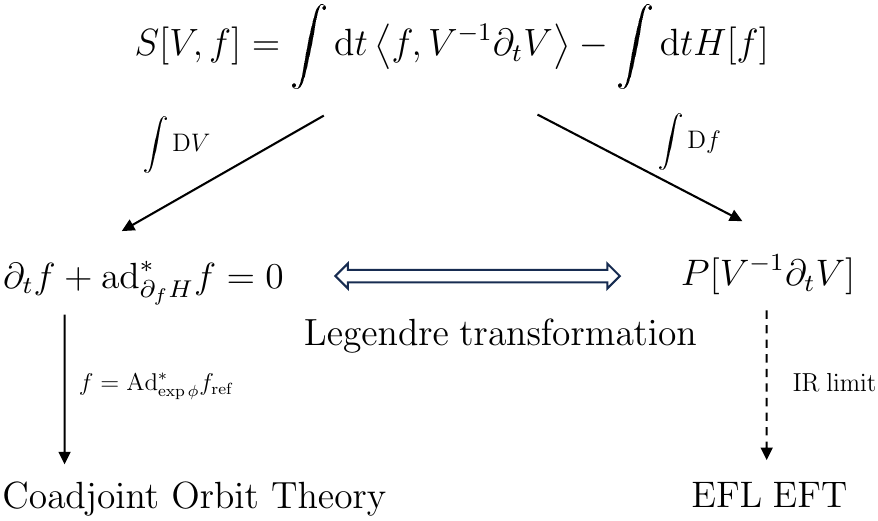}
\caption{The connection between the EFL EFT and the coadjoint orbit theory. }
\label{fig:legendre}
\end{figure}

In this section, we will show a concrete connection between the EFL EFT above, and a certain limit of the coadjoint orbit theory of  \cite{Delacretaz:2022coadj} (see \cite{Mehta2023postmodern} for a pedagogical review); see \figref{fig:legendre} for a summary of the result.

To begin, take the Lie group $\mG$ of canonical transformations of a single particle phase space $(\vec{x},\vec{p})$. Its Lie algebra, denoted by $\g$, is the space of all functions $F(\vec{x},\vec{p})$ which are interpreted as 1-particle observables. The dual space $\g^*$ is the space of 1-particle distributions $f(\vec{x},\vec{p})$. The Lie bracket in $\g$ is the Poisson bracket of functions $\{F,G\}$. The Lie algebra adjoint/coadjoint actions are given by $\mathrm{ad}_G F = \{G,F\}$ and $\mathrm{ad}^*_G f = \{G,f\}$, while the group adjoint/coadjoint actions are obtained by exponentiating the algebra adjoint/coadjoint actions
\begin{equation}
    \begin{split}
        \mathrm{Ad}_{\exp G} F &= e^{\mathrm{ad}_G} F = F + \{G,F\} + \ldots\\
        \mathrm{Ad}^*_{\exp G} f &= e^{\mathrm{ad}^*_G} f = f + \{G,f\} + \ldots\\
    \end{split}
\end{equation}
The group adjoint/coadjoint actions will often be denoted as conjugation, $\mathrm{Ad}_U F = U F U^{-1} $ and $\mathrm{Ad}^*_g f = g f g^{-1}$, for notational simplicity, even though the objects involved aren't matrices. Elements of the Lie group will be denoted by uppercase Latin letters starting from $U,V,\ldots$. Elements of the Lie algebra will be denoted by uppercase Latin letters starting from $F,G,\ldots$. Elements of the dual space will be denoted by lowercase Latin letters starting from $f,g,\ldots$. The exponential map from the Lie algebra to the Lie group will be denoted by $\exp$.

The most general action for a system whose configuration space is $\mG$ lives on the cotangent bundle $T^*\mG\cong \mG\times\g^*$ and is given by
\begin{align}\label{eq:S mechanics}
    S[V,f] = \int \ud t ~ \left\langle f ,V^{-1} \p_t V \right\rangle - \int \ud t ~ H[f],
\end{align}
where $\langle f,F \rangle = \int \ud^2 x \ud^2 p/(2\pi)^2 ~ Ff$ is the inner product between the Lie algebra and the dual space and $H$ is the Hamiltonian. We have demanded the action to be invariant under  left translation on $\mG$, i.e. the transformation $V\rightarrow WV$, so $H[f]$ cannot depend on $V$\footnote{Here we ignore the complication due to spontaneous symmetry breaking in which the Hamiltonian can depend on additional left-invariant vectors. }. Let us first vary the action with respect to $f$, and we get
\begin{align}\label{eq:g eom}
    V^{-1}\p_t V - \p_f H[f]=0.
\end{align}
Then, variation of $V$ is implemented by the transformation $V \rightarrow \exp(K)V$ for arbitrary $K(x,p)$, expanded to linear order in $K$, and we find $\delta S = \int \ud t \langle \pi,V^{-1}(\p_t K)V\rangle$. After some algebra, and using \eqref{eq:g eom}, we have
\begin{align}
    \p_t f = -\{ U^{-1}\p_t U,f \} = - \{ \p_f H, f \} = - \mathrm{ad}^*_{\p_f H} f.
\end{align}
This equation of motion on $\g^*$ is known as the Euler-Poisson equation \cite{khesin_2008_the}. One can show that time evolution as determined by this equation occurs via the action of canonical transformations on the initial state, so the phase space $\g^*$ get foliated into coadjoint orbits which don't mix under time evolution. If we restrict the dynamics to the coadjoint orbit which consists of all the points from $\mathrm{Ad}^*_{\exp \phi} f_{\mathrm{ref}}$ with a reference state $f_{\mathrm{ref}}$ and a field $\phi\in \g$, then there is a symplectic structure due to the Kirillov-Kostant-Souriau theorem \cite{aakirillov_2004_lectures}, from which we recover the coadjoint orbit action for $\phi$ studied in \cite{Delacretaz:2022coadj}.

Instead of integrating out $V$ from the action \eqref{eq:S mechanics}, we can equally integrate out $f$. Doing this in a saddle point approximation amounts to the Legendre transformation 
\begin{align}
    \left\langle f,V^{-1} \p_t V \right\rangle - H[f] = P[V^{-1} \p_t V],
\end{align}
where $P[\mu]$ is the pressure function for $\mu\equiv V^{-1} \p_t V$, the (phase space) chemical potential. Note that the pressure function is invariant under left-translations $V\rightarrow WV$ for any $W\in\mathcal{G}$. Let us denote $V = \exp \varphi,~\varphi\in \g$, and we will see that $\varphi$ corresponds to the dynamical field in the EFL EFT. The left translation symmetry of $V$ then turns into the shift symmetry $\varphi \rightarrow \varphi + \alpha + \frac{1}{2}\{ \varphi, \alpha \} + \ldots$ which, to leading order in both $\varphi$ and $\alpha$, is simply the hydrodynamic reparametrization symmetry \eqref{eq:reparametrization}. The nonlinear terms are a consequence of the fact that unlike $\mathrm{LU}(1)$, canonical transformations are nonabelian. Expanding the pressure in terms of its argument:
\begin{align}
    P[\mu] = P_0 + \left\langle \frac{\p P}{\p\mu}|_{\varphi=0} , \delta\mu  \right\rangle + \frac{1}{2} \left\langle \left\langle \frac{\p^2 P}{\p\mu \p \mu'}|_{\varphi=0} , \delta\mu'  \right\rangle , \delta\mu  \right\rangle + \frac{1}{3!} \left\langle \left\langle \left\langle \frac{\p^3 P}{\p\mu\p\mu' \p\mu''}|_{\varphi=0} , \delta\mu''  \right\rangle , \delta\mu'  \right\rangle , \delta\mu  \right\rangle +\ldots,
\end{align}
where $\p P/\p\mu\in \g^*$ given its differential forms and $P_0=P(\mu_0)$, $\delta\mu = \mu - \mu_0$. The chemical potential admits its own expansion in $\varphi$:
\begin{align}
    \mu = \mu_0 + \p_t \varphi + \frac{1}{2} \{ \p_t\varphi,\varphi \} + \frac{1}{6} \{ \{\p_t\varphi,\varphi\},\varphi\}+\ldots,
\end{align}
where the equilibrium value $\langle V^{-1}\p_t V\rangle_\text{eq} = \mu_0 $ is taken to be constant. 
We identify $\p P / \p \mu$ as the (phase space) density, or the distribution function $f$. Its deviation from the equilibrium distribution $f_0 = \Theta(p_\F - p) + \cdots$ can be related to the LU(1) charge density $\rho$ through the equation of motion (in the $q=0$ limit). Specifically, according to the perturbation \eqref{eq:Delta H}, we have $\p_t f = \{\delta H,f_0\} = - \vect n \cdot \vect E ~\delta(p-p_\F)$ where we used $\p_{p^i} f_0 = -\delta(p_\F - p) n^i$. Comparing it to \eqref{eq:anomalous conservation}, we find $f - f_0 = -C^{-1} \rho~ \delta(p-p_\F) $. Therefore, we can expand the distribution function as
\begin{align}
    f \equiv \frac{\p P}{\p\mu} = f_0 - \frac{\delta(p-p_\F)}{C} \int_{\theta'} g_2(\theta,\theta')\delta \mu(\theta') - \frac{\delta(p-p_\F)}{2C} \int_{\theta',\theta''} g_3(\theta,\theta',\theta'')\delta \mu(\theta')\delta \mu(\theta'')+\ldots,
\end{align}
Gathering the above, we arrive at 
\begin{align}\label{eq:P expansion}
    P[\mu] &= P_0 + \int \frac{\ud^2 x\ud^2 p}{(2\pi)^2} f_0(p) \left( \p_t \varphi + \frac{1}{2} \{\p_t\varphi,\varphi\}+ \frac{1}{6} \{\{\p_t\varphi,\varphi\},\varphi\}\right)(p)\nonumber\\
    &-\frac{p_\F}{2 C(2\pi)^2} \int \ud^2 x \ud\theta \ud\theta'~ g_2(\theta,\theta')\left(\p_t \varphi(\theta)\p_t \varphi(\theta')+ \frac{1}{2}\p_t \varphi(\theta) \{\p_t\varphi,\varphi\}(\theta') + \frac{1}{2}\p_t \varphi(\theta') \{\p_t\varphi,\varphi\}(\theta) \right)\nonumber\\
    & - \frac{p_\F}{6 C(2\pi)^2} \int \ud^2 x \ud\theta \ud\theta' \ud\theta''~ g_3(\theta,\theta',\theta'') ~\p_t \varphi (\theta) \p_t \varphi (\theta')  \p_t \varphi (\theta'')+ O(\varphi^4).
\end{align}
The term linear in $\varphi$ is a total derivative, thus we ignore it. At the quadratic level, we have
\begin{align}\label{eq:P 2}
    P^{(2)} =  -\frac{p_\F}{2 C(2\pi)^2} \int \ud^2 x \ud\theta \ud\theta'~ g_2(\theta,\theta') \p_t \varphi(\theta)\p_t \varphi(\theta') - \frac{p_\F}{2(2\pi)^2} \int \ud^2 x \ud\theta ~ \vect n \cdot\nabla\varphi \p_t \varphi.
\end{align}
By comparing to the EFL EFT action \eqref{eq:S 2}, we arrive at
\begin{align}
    C = - \frac{p_\F}{(2\pi)^2}.
\end{align}
With this value of $C$, we see that the commutation relation \eqref{eq:KM algebra} agrees with that in \cite{Delacretaz:2022coadj}, and the Luttinger's theorem reads $\nu = \mathcal{V}_\F/(2\pi)^2$.

The most general cubic pressure $P^{(3)}$ does not match the cubic EFT action $S^{(3)}$ in \eqref{eq:S 3} due to the additional cross terms upon expanding the pressure function. However, in the long wanvelength limit $q = 0$, the cross terms containing the Poisson bracket vanish and only the third line in \eqref{eq:P expansion} survives. Hence, we see that $\int \ud t~ P^{(3)}|_{q=0} = S^{(3)}.$

The only discrepency between our EFL EFT and the coadjoint orbit action appears at finite $q$. This can be intuitively understood as that LU(1) is a larger symmetry that forbids finite-$q$ transfer among different points on the Fermi surface. More precisely, the reparametrization symmetry that protects our EFL EFT from having $q$-dependent terms is broken explicitly in the coadjoint orbit method. Indeed, in the $q=0$ limit, the action \eqref{eq:S mechanics} only contains $\p_t \varphi$, so after integrating out $f$, the resulting Lagrangian formalism can only be functions of $\p_t \varphi$.  

To summarize, our EFL EFT is the Legendre transformation of the coadjoint orbit action for Fermi liquid up to irrelevant, LU(1)-breaking corrections including cubic and higher order terms in the dynamical field, and the nonlinear part of the action agrees with the coadjoint orbit method in the $q=0$ limit.

\section{Towards non-Fermi liquids?}\label{sec:nfl}

We have argued above that the EFL EFT by itself reproduces the EFT of a conventional Fermi liquid.  Here, let us remark on one possible route to a low-energy EFT that manifestly looks like a non-Fermi liquid, while enforcing LU(1) symmetry.  Suppose that the charge density of LU(1) is coupled to an additional bosonic degree of freedom $\Phi$. The ``minimal'' coupling that obeys LU(1) symmetry is given by
\begin{align}\label{eq:Sint}
    S_{\mathrm{int}} = \int \ud t \ud^2 x \ud\theta ~ \lambda(\theta) \Phi(x^\mu) \p_t \varphi(x^\mu,\theta).
\end{align}
One can generalize the boson field to be multicomponent, but, for simplicity, we focus on a single boson mode. Such an interaction describes the coupling between different points on the Fermi surface mediated by the bosons, and coincides with the ``mid-IR'' theory studied in \cite{Senthil_gift, shi2023loop}. The full effective action is then supplemented by actions for $\varphi$ and $\Phi$: $S = S_\varphi+S_{\mathrm{int}}+S_\Phi$, where $S_\varphi$ is the LU(1) action and we can parametrize $S_\Phi$ by
\begin{align}
    S_\Phi = \frac{1}{2} \int \ud t \ud^2 x ~\left(\p_t^{\frac{\epsilon}{z}} \Phi\right)^2 - m_c^2 \Phi^2 - J\left(\nabla^\epsilon \Phi\right)^2 + \cdots,
\end{align}
where $\cdots$ denote nonlinear corrections and  $m_c^2$ is tuned to a critical point (the value of which depends on regularization scheme). \textcolor{red}{Note that the critical exponents $z,\epsilon$ could a priori take nontrivial values determined by the IR fixed point; we will not attempt to derive such $z$ and $\epsilon$ self-consistently in this paper, but rather explore the consequences of these exponents.}
If we treat the Gaussian theory above as exact, and integrate out the boson $\Phi$, we will obtain a new quadratic effective action (at leading order) for $\partial_t \varphi$: \begin{equation} \label{eq:nonlocal}
    \mathrm{\Delta}S_{\mathrm{EFL}} \sim -\lambda^2 \int \mathrm{d}\omega \mathrm{d}q \mathrm{d}\theta \mathrm{d}\theta^\prime \frac{\partial_t \varphi(\theta; \omega,q)\partial_t \varphi(\theta^\prime; -\omega,-q)}{\omega^{2\epsilon/z} - q^{2\epsilon}}.
\end{equation}
Notice that such a theory will continue to have LU(1) symmetry; however, it is no longer spatially local.  \textcolor{red}{Note that the bosonic action already is generally nonlocal for noninteger $z$ or $\epsilon$.}  The framework developed in this paper, based simply on the LU(1) symmetry, is not enough to systematically deduce the most general such non-local theory.  Non-local couplings such as (\ref{eq:nonlocal}) generally will qualitatively modify the zero sound dispersion, at a minimum.  

In \cite{Kim1995qbe,Else_2023}, the authors find that the effective bosonic action does not effectively have a critical mass: integrating out $\Phi$ above simply modifies the Landau parameter $F_0$.\footnote{Strictly speaking, the Boltzmann equation is valid for the first few angular momentum modes where the singular self-energy behavior will be exactly cancelled by a singular Landau interaction \cite{Kim1995qbe}.  This, however, should not be regarded as a contradiction to our conclusion since, by assumption, a local EFT does not permit any singular Landau interaction.} Such a case is already handled by the effective field theory of Section \ref{sec:eft}. We emphasize that any non-local effective action (\ref{eq:nonlocal}) requires fine-tuning which may be unphysical, but such fine-tuning is needed to avoid the conclusion of Section \ref{sec:eft} that the EFL EFT appears identical to the FL EFT up to irrelevant corrections.

Lastly, we remark that some of our other scaling assumptions, including that $[\theta] = 0$, could be relaxed in order to provide another route to NFL physics.  However, in order to avoid an infinite tower of increasingly relevant interactions, additional structure must be imposed to constrain the Lagrangian.\footnote{This is the case for the coadjoint orbit theory coupled to a critical boson. While here one also has an infinite tower of relevant terms with $[\theta]>0$ (see section VI.A of Ref.\ \cite{Mehta2023postmodern}), the tower of relevant terms there is very strongly constrained, and there are only a finite number of independent coefficients for the relevant terms.  This allows for more control than we have in the EFL EFT.}  In the EFL formalism, this inevitably requires \emph{further} symmetries or constraints beyond LU(1) symmetry.

\section{Adding dissipation}
To construct a dissipative EFT at finite temperature, we put the action on the Schwinger-Keldysh (SK) contour with two legs labeled by $s=1,2$; see recent development of dissipative EFT \cite{Crossley:2015evo,Glorioso:2017fpd}, which we will review only tersely here.  It is convenient to work with the $r,a$-variables defined as
\begin{align}
    \Psi_r = \frac{\Psi_1+\Psi_2}{2},\quad \Psi_a = \Psi_1 - \Psi_2,
\end{align}
where $\Psi_s$ denote collectively the background and dynamical fields. Most of the symmetries in \secref{sec:symmetry} generalize straightforwardly to the two copies, except for the reparameterization symmetry which becomes ``diagonal''
\begin{align}
    \varphi_s\to \varphi_s+\chi(x^i,\theta)
\end{align}
with the \emph{same} shift on the SK contour. Therefore, the simplest invariant building blocks are
\begin{align}
    \mu\equiv B_{r,t},\quad B_{a,\mu}.
\end{align}
The presence of two copies of each field implies that the resulting effective field theory will combine both dissipative effects as well as stochastic fluctuations; the presence of both is mandated by the fluctuation-dissipation theorem.

In order to describe a system approaching thermal equilibrium, the generating function should satisfy an additional $\mathbb{Z}_2$ symmetry, known as the Kubo–Martin–Schwinger (KMS) symmetry. (This is the generalization of time-reversal symmetry to the dissipative effective field theory.) Consider a system that preserves $\mathcal{T}\mathcal{I}$, where $\mathcal{T}$ is the time reversal and $\mathcal{I}$ is the inversion. The KMS transformation in the classical limit is given by 
\begin{subequations}\label{eq:KMS}
    \begin{align}
        \widetilde B_{r,\mu}(-x^\mu,\theta) &=  B_{r,\mu}(x^\mu,\theta),\\
        \widetilde B_{a,\mu}(-x^\mu,\theta) &= B_{a,\mu}(x^\mu,\theta)  +\ii \beta \p_t B_{r,\mu}(x^\mu,\theta),\\
        \tilde \varphi_r(-x^\mu,\theta) &= -\varphi_r(x^\mu,\theta),\\
        \tilde \varphi_a(-x^\mu,\theta) &= -\varphi_a(x^\mu,\theta)-\ii\beta \p_t \varphi_r(x^\mu,\theta),
    \end{align}
\end{subequations}
where $\beta$ is the inverse temperature, and $\theta$ is taken to be $\mathcal{T}\mathcal{I}$-even.

The most general first-order dissipative EFT can be written as
\begin{align}\label{eq:S dissipative}
    S &= \int \ud t \ud^2 x \ud\theta \ud \theta'~\left[ g_2(\theta,\theta') B_{r,t}(x^\mu,\theta) B_{a,t}(x^\mu,\theta') + \mathrm{i} \left(n^i \;\; s^i\right) \Sigma(\theta,\theta') \left(\begin{array}{c} n'^j  \\ s'^j \end{array}\right) B_{a,i}(x^\mu,\theta)\left( B_{a,j} + \ii\beta \p_t B_{r,j}\right)(x^\mu,\theta')   \right]\nonumber\\
    & \quad + \frac{C}{2}\int \ud t \ud^2 x \ud\theta ~\left[ n^i B_{a,i}B_{r,t} + n^i B_{r,i}B_{a,t} + \varphi_a F_{r,  i t} n^i + \varphi_r F_{a, i t} n^i  \right],
\end{align}
where $s^i = \epsilon^{ji}n_j$ and $\Sigma$ is a $2\times 2$ symmetric positive-definite noise matrix.\footnote{Notice that we only require $\Sigma_{12}(\theta',\theta) = \Sigma_{21}(\theta,\theta') $, so $\Sigma_{12} = \pm \Sigma_{21}$ for them to be even/odd functions of $\theta-\theta'$ assuming rotational symmetry.} The first line in \eqref{eq:S dissipative} characterizes the equilibrium pressure (the first term) and the first-order dissipation and fluctuation (the second term), while the second line generates the anomaly \eqref{eq:W anomaly}. Explicitly, by varying with respect to $\varphi_a$ and then turning off all the $a$-variables, we obtain the dissipative, noise-free equation of motion
\begin{align}\label{eq:eom diss}
    \p_t \rho(x^\mu,\theta) - \beta\int \ud\theta' \Sigma^{ij}(\theta,\theta')\p_i\p_j \delta \mu(x^\mu,\theta')  + C\vn \cdot\nabla \delta \mu(x^\mu,\theta) = C\vn \cdot \vect E - \beta \int \ud\theta' \Sigma^{ij}(\theta,\theta') \p_i E_j,
\end{align}
where we denoted $\Sigma^{ij}(\theta,\theta')\equiv \left(n^i \;\; s^i\right) \Sigma(\theta,\theta') \left(n'^j \;\; s'^j\right)^{\mathrm{T}}$.
In the non-interacting limit $\delta\mu =C^{-1} v_\F \rho$ and diagonal noise matrix $\Sigma^{ij}(\theta,\theta') = C D v_\F^{-1} ~\delta^{ij}\delta(\theta-\theta')$, the quasinormal modes have \begin{equation}
    \omega = v_\F \vect n \cdot \vect q - \ii D q^2 \label{eq:diffusive}
\end{equation}
which features a Fermi surface chiral mode damped by diagonal U(1) diffusion. The normal mode structure could be more complex if $g_2$ and $\Sigma^{ij}$ are non-trivial functions, but the quadratic scaling of the dissipative contribution will persist.  In general, we will have a dynamical scaling exponent $z\neq 1$ in the dissipative EFL EFT.

We can now repeat the same dimensional analysis of Section \ref{sec:eft}, now deducing if the \emph{dynamical universality class} of EFL hydrodynamics is stable or not from an RG perspective.  Such a possible instability would mimic the instability of the Navier-Stokes equations \cite{spohn_nonlinear_2014}, or biased diffusion \cite{Delacretaz_2020}, in one spatial dimension  to the Kardar-Parisi-Zhang fixed point. Repeating the scaling analysis from before using $[q_\parallel]=1$, $[q_\perp] = \alpha$ and $[\omega]=z$, we find that \begin{equation}
    z = \min(2,2\alpha),\;\;\;\; [\varphi_a] = [\partial_t\varphi_r] = \frac{1+\alpha}{2}.
\end{equation}
Notice that if $\alpha\ne 1$, one of the components of noise is irrelevant.\footnote{If $\alpha \ne 1$, then naively both (\ref{eq:diffusive}) needs drastic corrections to the decay rate of the hydrodynamic modes, yet there are also no relevant operators capable of doing so (at least without coupling to additional slow degrees of freedom).  From this perspective, one may require $\alpha=1$ within the dissipative EFT, and/or only allow diffusion in the $n_i$ direction.}  The most relevant nonlinearity in this problem is subtle \cite{Delacretaz_2020} and is easiest to see by changing variables from $\mu = \partial_t \varphi_r$ to the density $\rho$ using thermodynamic relations \eqref{eq:noether}.  In this case, the quadratic nonlinearity in the thermodynamic relation $\mu(\rho)$ gives rise to a cubic nonlinearity in the anomaly term of the action above, whose scaling dimension can be shown to be non-negative in any $d\ge 2$.  We thus deduce that, up to the likely marginal relevant corrections in $d=2$ which are analogous to the marginally relevant corrections to the Gaussian fixed point of the Navier-Stokes equations  in $d=2$ \cite{dorfman}, the LU(1) hydrodynamics is stable as a dynamical fixed point as well.

\section{Conclusion}

We have presented above the general hydrodynamic effective field theory of an LU(1)-symmetric system, i.e. an EFL. Our analysis leads us to conclude that as long as the LU(1) densities are the only low energy degree of freedom and one can build a local EFT for the EFL, the zero temperature thermodynamics \emph{and dynamics} of an EFL look identical to an ordinary Fermi liquid in the IR, differing only in irrelevant corrections. Similar conclusions hold for the \emph{dynamical} fixed point of LU(1) hydrodynamics at finite temperature, up to possibly marginally relevant perturbations in $d=2$.

The renormalization group in the presence of a Fermi surface is quite subtle in general, and while we have ignored some of these subtleties in our analysis, such as potential UV/IR mixing in loops as well the question of nonzero scaling dimension for the angle $[\theta]\neq 0$, our goal here is not to solve these hard questions but instead to address the question of whether LU(1) as an emergent symmetry can circumvent such concerns and ultimately lead to a better-posed EFT for non-Fermi liquids. We found that the Gaussian truncation of the EFL EFT is equivalent to earlier approaches to higher-dimensional bosonization of Fermi surfaces \cite{haldane2005luttingers, CastroNetoFradkin:1994, Houghton:2000bn}, albeit without the need for a patch decomposition of the Fermi surface.  As we showed, the LU(1) approach is very similar to the coadjoint orbit theory \cite{Delacretaz:2022coadj,Mehta2023postmodern} and differs only in irrelevant corrections. As such, the EFL EFT provides a systematic prescription for nonlinear corrections to higher dimensional bosonization, albeit one that differs from the coadjoint orbit theory.

Since the degree of freedom in the EFL EFT is inherently bosonic, this formalism can also capture Luttinger liquids in 1d, which are non-Fermi liquids in the sense of having the microscopic fermion destroyed at low energies. From a low energy perspective, however, an effective description does not readily identify such details.  We therefore leave open the possibility that an EFL EFT describes non-Fermi liquids where the \emph{fermionic} Green's functions are qualitatively changed, albeit the bosonic collective modes appear identical to an ordinary Fermi liquid.  However, we still expect that a general non-Fermi liquid phase could be distinguished from conventional metallic phases via bosonic observables (electrical conductivity, plasmon dispersion, thermodynamic susceptibilities, etc.), and such bosonic observables are within the purview of our EFT. From this perspective, the simplest EFLs look (at leading order) identical to Fermi liquids at a first glance.

\section*{Acknowledgements}
We acknowledge useful discussions with Dominic Else, Hart Goldman, and Cenke Xu.
This work was supported by the U.S. Department of Energy, Office of Science, Basic Energy Sciences under Award number DESC0014415 (MQ), the Alfred P. Sloan Foundation through Grant FG-2020-13795 (UM, AL), and the Gordon and Betty Moore Foundation's EPiQS Initiative via Grant GBMF10279 (XH, AL). XH thanks the Kavli Institute for Theoretical Physics for hospitality; KITP is supported in part by the Heising-Simons Foundation, the Simons Foundation, and the National Science Foundation under Grant PHY-2309135.

\appendix

\bibliography{lu1}

\end{document}